\documentclass[prl,twocolumn,
 amsmath,amssymb,
 aps,showpacs,superscriptaddress
]{revtex4-1}

\usepackage{graphicx}% Include figure files
\usepackage{dcolumn}% Align table columns on decimal point
\usepackage{bm}% bold math
\usepackage{subfig}
\usepackage{color}

\DeclareCaptionJustification{justified}{\leftskip=0pt \rightskip=0pt \parfillskip=0pt plus 1fil}
\newcommand{\beqar}{\begin{eqnarray}}
\newcommand{\eeqar}{\end{eqnarray}}

\newcommand{\bcen}{\begin{center}}
\newcommand{\ecen}{\end{center}}

\newcommand{\f}[2]{\frac{#1}{#2}}
\renewcommand{\b}[1]{\left({#1}\right)}
\renewcommand{\v}[1]{\vec{#1}}
\newcommand{\pd}[2]{\frac {\partial #1}{\partial #2}}

\renewcommand{\sb}[1]{\left[{#1}\right]}
\newcommand{\mean}[1]{\langle {#1} \rangle}

\newcommand{\ra}{\rightarrow}

\newcommand{\da}{\downarrow}
\newcommand{\ua}{\uparrow}

\begin{document}

\preprint{APS/123-QED}
\title{Shortcut to Equilibration of an Open Quantum System}
%\title{Induced Swift Equilibration for an Open Quantum System}% Force line breaks with \\

\author{Roie Dann}
\email{roie.dann@mail.huji.ac.il}
%\altaffiliation[Also at ]{The Institute of Chemistry, The Hebrew University of Jerusalem, Jerusalem 9190401, Israel.}%Lines break automatically or can be forced with \\
%\altaffiliation{Kavli Institute for Theoretical Physics, University of California, Santa Barbara, CA 93106, USA}
\affiliation{The Institute of Chemistry, The Hebrew University of Jerusalem, Jerusalem 9190401, Israel}%
\affiliation{Kavli Institute for Theoretical Physics, University of California, Santa Barbara, CA 93106, USA}
\author{Ander Tobalina}
\email{ander.tobalina@ehu.eus}
\affiliation{Department of Physical Chemistry, University of the Basque Country UPV/EHU, Apdo 644, Bilbao, Spain}
\affiliation{Kavli Institute for Theoretical Physics, University of California, Santa Barbara, CA 93106, USA}
\author{Ronnie Kosloff}%
\email{kosloff1948@gmail.com}
\affiliation{The Institute of Chemistry, The Hebrew University of Jerusalem, Jerusalem 9190401, Israel}%
\affiliation{Kavli Institute for Theoretical Physics, University of California, Santa Barbara, CA 93106, USA}

\date{\today}% It is always \today, today,
\begin{abstract}
    We present a procedure to accelerate the relaxation of an open quantum system towards its equilibrium state. The control protocol, termed {\emph{Shortcut to Equilibration}}, is obtained by reverse-engineering the non-adiabatic master equation. 
    This is a non-unitary control task aimed at rapidly changing the entropy of the system.
    Such a protocol serves as a shortcut to an abrupt change in the Hamiltonian, i.e., a {\em{quench}}. As an example, we study the thermalization of a particle in a harmonic well. We observe that for short protocols there is a three orders of magnitude improvement in accuracy.
\end{abstract}

%The pacs numbers mean: quantum mechanics, decohernece: open system, Coherent control of atomic interactions with photons, algebriac methods
% CHECK IF THE LAST ONE IS RELEVANT                         
\maketitle

 \paragraph*{Introduction}
 \label{sec:intro}
Equilibration is a natural process, describing the return of a perturbed system back to a thermal state.
%at a typical relaxation rate $\tau_R^{-1}$.
The relaxation to equilibrium is present in both the classical \cite{thermodynamics1956fermi,hecht1990statistical,martinez2016engineered} and quantum \cite{breuer2002theory} regimes. Gaining control over the relaxation rate of quantum systems is crucial for enhancing the performance of quantum heat devices \cite{alicki1979quantum,kosloff2017quantum,geva1992quantum,feldmann2003quantum,dambach2018quantum}. In addition, fast relaxation is beneficial for quantum state preparation \cite{verstraete2009quantum,ye2008quantum} and open system control \cite{koch2016controlling,blaquiere1987information,huang1983controllability,brockett1983differential,d2007introduction,brif2010control}. 
To address these issues, we present a scheme to accelerate the equilibration of an open quantum system, serving as a shortcut to the natural relaxation time $\tau_R$. The protocol is termed {\emph{Shortcut To Equilibration}} (STE).

This control problem is embedded in the theory of open quantum systems \cite{breuer2002theory}. The framework of the theory assumes a composite system, partitioned into a system and an external bath. The Hamiltonian describing the evolution of the composite system reads $\hat{H}\b t=\hat{H}_S\b t+\hat{H}_B+\hat{H}_I$, where $\hat{H}_S\b t$ is the system Hamiltonian, $\hat{H}_B$  is the bath Hamiltonian and $\hat{H}_I$ is the system-bath interaction term. When the system depends explicitly on time, the driving protocol influences the system-bath coupling operators and consequently, the relaxation time.

Quantum control in open systems has been addressed in the past utilizing measurement and feedback \cite{lloyd2000control,viola1999dynamical,liu2011experimental,khodjasteh2010arbitrarily,schmidt2011optimal,altafini2004coherent}.
Typically, the effect of non-adiabatic driving on the dissipative dynamics was ignored  \cite{vacanti2014transitionless,suri2017speeding,jing2013inverse,scandi2018thermodynamic}. Here, we present a comprehensive theory that incorporates the non-adiabatic effects. The formalism is based on the recent derivation of the Non Adiabatic Master Equation (NAME) \cite{dann2018time}. This master equation is of the Gorini-Kossakowski-Lindblad-Sudarshan (GKLS) form, guaranteeing a complete positive trace-preserving dynamical map \cite{lindblad1976generators,gorini1976completely,alicki2007general}. A further prerequisite is the {\em{inertial theorem}} \cite{dann2018inertial}. This theorem allows extending the validity of the NAME for processes with small  `acceleration' of the external driving.

\begin{figure}[htb!]
\centering
\includegraphics[width=8cm]{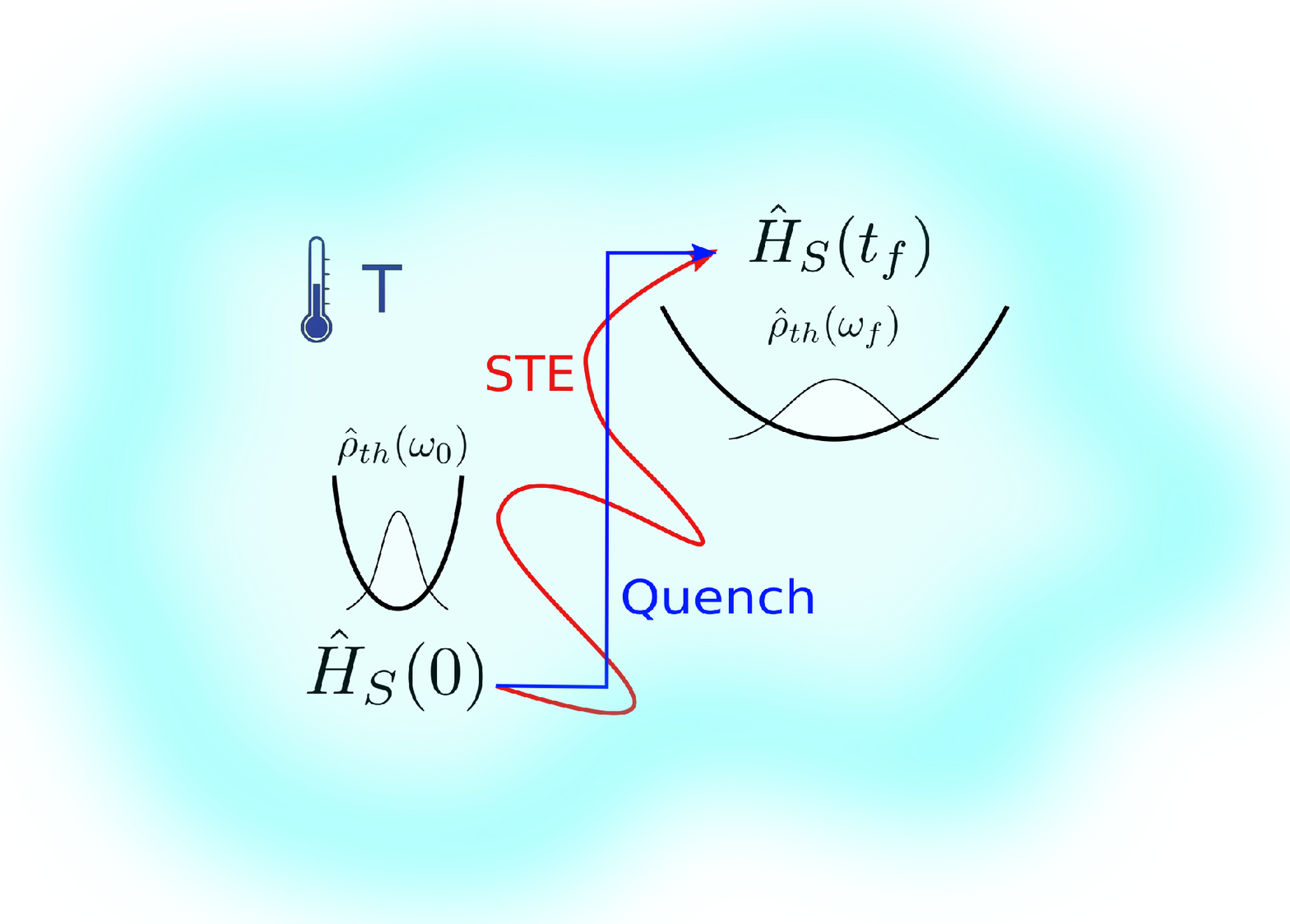}
\caption{Scheme of the {\em{Shortcut To Equilibration}} (STE) protocol (curved red line) and the {\em{quench}} protocol (blue step line), transforming an initial thermal state at temperature $T$ and frequency $\omega_i$ to a final thermal state with an equivalent temperature and frequency $\omega_f$.}
\label{fig:scheme}
\end{figure}

We consider a driven quantum system, the Hamiltonian of which varies from $\hat{H}_S\b 0$ to a final Hamiltonian $\hat H_S\b {t_f}$, while coupled to a thermal bath (see Fig. \ref{fig:scheme}). Our aim is to exploit the non-adiabatic effects of the driving to accelerate the system’s return to equilibrium. By  reverse-engineering the NAME, we find a protocol that transforms the thermal state of $\hat{H}_S\b 0$ at temperature $T$ to the corresponding thermal state of $\hat H_S\b {t_f}$. This procedure serves as a shortcut for the natural relaxation time $\tau_R$. 

Controlling the equilibration rate differs from the control tasks treated by shortcuts to adiabaticity \cite{demirplak2003adiabatic,berry2009transitionless,chen2010fast,muga2010transitionless,stefanatos2010frictionless,hoffmann2011time,del2013shortcuts,torrontegui2013shortcuts,abah2017performance,reichle2006transport}. The latter protocols generate an entropy-preserving unitary transformation, which is effectively the identity map between initial and final diagonal states in the energy representation. 
Conversely, the STE procedure is a non-unitary transformation, which is designed to rapidly change the entropy of the system. 

%\paragraph*{Shortcut to equilibration}
\paragraph*{System dynamics}
We consider a quantum particle in contact with a thermal bath while confined by a time-dependent harmonic trap. The system Hamiltonian reads 
 \begin{equation}
 \hat{H}_S \b t=\f{\hat{P}^2}{2m}+\f{1}{2}m \omega^2\b t \hat{Q}^2~,
 \label{eq:HO Ham}
 \end{equation}
 where $\hat{Q}$ and $\hat{P}$ are the position and momentum operators, respectively, $m$ is the particle mass and $\omega \b t$ is the time-dependent oscillator frequency. We assume a Bosonic bath with 1D Ohmic spectral density and an interaction Hamiltonian of the form $\hat{H}_I= -\hat{D}\otimes\hat{B}$, where $\hat{D}=d\hat{a}+d^*\hat{a}^{\dagger}$ ($a$ and $a^{\dagger }$ are the annihilation and creation operators of the oscillator, respectively), $d$ is the interaction strength and  $\hat{B}$ is the bath interaction operator.
Throughout the paper, we choose units related to the minimum frequency $\omega_{min}$, time $2\pi/\omega_{min}$ and energy $\hbar \omega_{min}$ with $\hbar=1$. 

At initial time, the open quantum system is in  equilibrium with the bath, and the state is of a Gibbs canonical form  $\hat{\rho}_S \b 0=Z^{-1}e^{-\hat{H}_S\b 0/k_B T}$, where $Z$ is the partition function, $k_B$ is the Boltzmann constant and $T$ is the temperature of the bath. We search for a protocol that varies the Hamiltonian toward $\hat{H}_S \b {t_f}$ with a target thermal state $\hat{\rho}_{S}^{Th} \b{t_f}=Z^{-1}e^{-\hat{H}_S\b{t_f}/k_B T}$. This procedure serves as a shortcut to an isothermal process.
The accuracy of this transformation can be quantified using the fidelity $\cal{F}$, which is a measure of the distance between the final state $\hat{\rho}_S \b{t_f}$ of the protocol and $\hat{\rho}_{S}^{Th} \b{t_f}$ \cite{isar2009quantum,scutaru1998fidelity,banchi2015quantum}. A classical analogous problem has been addressed by Martinez et al. \cite{martinez2016engineered}.

The most straightforward protocol is a  {\emph{quench}} protocol. 'Quench' means abruptly changing the Hamiltonian from $H_S\b 0$ to $H_S\b{t_f}$, and then letting the system equilibrate with the bath, Cf. Supplemental Material (SM) III.
 When  $\hat{H}_S\b 0$ and $\hat{H}_S \b{t_f}$ do not commute, which is the case for a non-rigid harmonic oscillator, such a sudden change generates coherence in the energy basis, leading to deviations from equilibrium. 
 The quenched system relaxes at an exponential rate toward equilibrium, which leads to an asymptotic exponential convergence of the fidelity toward unity $1-{\cal{F}}\b t\propto e^{-k t}$, for $t/k>1$, with $k=k_\da-k_\ua$, where $k_\da$ and $k_\ua$ are %the constant relaxation and excitation 
 decay rates, Cf. SM IIIA.
 We use the {\emph{quench}} protocol as a benchmark to assess the STE protocol's performance. 

To describe the reduced dynamics under the STE, we follow the derivation presented in Refs. \cite{dann2018time,dann2018inertial}. First, we obtain a solution for the unitary propagator $\hat{U}_S\b{t,0}$
for a protocol determined by a constant adiabatic parameter $\mu=\dot{\omega}/\omega^2$. The closed-form solution of $\hat{U}_S\b{t,0}$ allows constructing a master equation that includes the bath's influence on the reduced dynamics.  Then, by utilizing the {\em{inertial theorem}}, we extend the description to protocols where $\mu$ varies slowly ($d\mu/dt\ll1$). This condition sets a lower bound for the minimum protocol duration. For protocols faster than the minimum time, the condition $d\mu/dt\ll1$ is no longer satisfied and the inertial approximation loses its validity \cite{dann2018inertial}. The bound is given by $t_{f}>f\cdot\text{max}_{s}\b{\f{1}{\omega}\sqrt{\f{\omega''\b{s}}{2\omega}\f 1{8-\mu^{2}}}}$, where $s=t/t_f$ and $f<1$ is a small scalar, dependent on the desired precision, Cf. SM II. For example, if $f=0.05$, the lower bound is $t_f> 4.38\,\, \b{2\pi/\omega_{min}}$, where $\omega_{min}=5\, \text{a.u}$. 

%where $U\b{t,0}$ is obtained within a closed Lie algebra \cite{dann2018inertial}, Cf. Appendix \tb{\ref{}}
The range of validity of the NAME sets a number of conditions: (i) weak coupling between system and bath, which also allows for a reduced description of the system's dynamics in terms of $\hat{\rho}_S$ \cite{breuer2002theory}; 
(ii) Markovianity \footnote{Markovianity, as used in this paper, is the condition of timescale separation between a fast decay of the bath correlations and the slow system dynamics.}; 
(iii) large Bohr frequencies relative to the relaxation rate $\tau_R$; 
(iv) slow driving relative to the decay of the bath correlations.
In the following, we consider a regime where the NAME and {\em{inertial theorem}} are valid.
%A restriction arises due to the validity of the inertial approximation; (5) Slow acceleration of the driving. For a certain protocol, this condition determines a lower bound to the minimal protocol duration. For protocols faster than the minimum time, the condition $d\mu/dt\ll1$ is no longer satisfied. The bound is given by $t_{f}>f\cdot\text{max}_{s}\b{\f{1}{\omega}\sqrt{\f{\omega''\b{s}}{2\omega}\f 1{8-\mu^{2}}}}$, where $s=t/t_f$ and $f<1$ is a small scalar, dependent on the desired precision, Cf. Appendix \ref{sec:lower bound}. For $f=0.05$ the lower bound is $t_f> 5.5\, \text{a.u}$.

The dynamics of the externally driven open quantum system, in the interaction representation, is described by
\begin{multline}
\f d{dt}\tilde{\rho}_{S} \b t=k_{\downarrow} \b t\b{\hat{b}\tilde{\rho}_{S}\b t\hat{b}^{\dagger}-\f 12\{\hat{b}^{\dagger}\hat{b},\tilde{\rho}_{S}\b t\}}\\
+k_{\uparrow}\b t \b{\hat{b}^{\dagger}\tilde{\rho}_{S}\b t\hat{b}-\f 12\{\hat{b}\hat{b}^{\dagger},\tilde{\rho}_{S} \b t\}}~~~.
\label{eq:NAME}
\end{multline}
Here, the interaction picture density operator reads $\tilde{\rho}_S\b{t}=\hat{U}\b{t,0}\hat{\rho}_S \b t \hat{U}^{\dagger}\b{t,0} \label{eq:int} $. We use the notation $\tilde{A}$ to describe operators in an interaction picture relative to the system Hamiltonian.
For an ohmic Bosonic bath, the decay rates are 
\begin{equation}
k_{\da}\b t=k_{\ua} \b t e^{\alpha\b t/k_B T}=\f{\alpha \b t|\v d|^{2}}{8\pi \varepsilon_{0} \hbar c}\b{1+N\b{\alpha\b t}}~,
\label{eq:k down}
\end{equation}
where $N$ is the occupation number of the Bose-Einstein distribution and $\alpha$ is a modified frequency, determined by the non-adiabatic driving protocol \cite{dann2018time}. In terms of the oscillator frequency, the modified frequency is given by
\begin{equation}
\alpha \b t= \sqrt{1-\f{1}{4}\b{\f{\dot{\omega}\b t}{\omega^2 \b t}}^2} \omega \b t~~.    
\label{eq:alpha1}
\end{equation}
The Lindblad jump operators become $\hat{b}\equiv \hat{b} \b 0 =\sqrt{\f{m\omega\b 0}{2\hbar}}\f{\left(\kappa+i\mu\right)}{\kappa}\b{\hat{Q}\b{0}+\f{\mu+i\kappa}{2m\omega\b 0}\hat{P}\b{0}}$ where $\kappa = \sqrt{4-\mu^2}$. 

In the interaction representation the Lindblad operators are time-independent. This property provides an explicit solution in terms of the second-order moments ${\cal{B}}=\{\hat{b}^\dagger \hat{b},\hat{b}^2,\hat{b}^{\dagger2}\}$ \cite{dann2018time,dann2018inertial}, Cf. SM I, which, together with the identity operator, form a closed Lie algebra. The solution is given by a generalized canonical state, which has a Gaussian form in terms of $\cal{B}$. Such states are canonical invariant under the dynamics described by Eq. \eqref{eq:NAME}, implying that the system can be described by the generalized canonical state throughout the entire evolution \cite{alhassid1978connection,Jaynes1957,rezek2006irreversible,andersen1964exact}. The system state is given by
\begin{equation}
  \tilde{\rho}_S \b t= Z^{-1}e^{\gamma\b t \tilde{b}^2}e^{\beta\b t \tilde{b}^{\dagger}\tilde{b}}e^{\gamma^{*}\b t \tilde{b}^{\dagger 2}}~~,  
  \label{eq:Gibbs state}
\end{equation}
  which is completely defined by the time-dependent coefficients $\gamma$ and $\beta$ and the driving protocol. The partition function reads $Z\b{\beta,\gamma}=\f{e^{-\beta}}{\b{e^{-\beta}-1}\sqrt{1-4|\gamma|^{2}/\b{e^{-\beta}-1}^{2}}}$. 
In the adiabatic limit, the adiabatic parameter $\mu$ approaches zero, the state follows the adiabatic solution, and $\hat{b}^{\dagger}\hat{b}\ra \hat{a}^{\dagger}\hat{a}$.

Substituting $\tilde{\rho}_S\b t$ into the master equation, Eq. \eqref{eq:NAME}, multiplying by $\tilde{\rho}_S^{-1}$ from the right and comparing the terms proportionate to the operators  $\tilde{b}^\dagger \tilde{b}$, $\tilde{b}^2$ and $\tilde{b}^{\dagger 2}$  leads to
\begin{eqnarray}
\begin{array}{l}
\dot{\beta}=k_{\da}\b{e^\beta -1}+k_{\ua}\b{e^{-\beta}-1+4 e^{\beta}|\gamma|^2},\\
 \dot{\gamma}= \b{k_\da+k_\ua}\gamma-2 k_{\da} \gamma e^{-\beta}~~.
\end{array}
\label{eq:beta_gamma}
\end{eqnarray}
These equations describe the evolution of the system for any initial squeezed thermal state.
Here, we assume that the system is in a thermal state at the initial time, which infers $\gamma(0)=0$. This simplifies the expression of the state to
\begin{equation}
    \tilde{\rho}_S \b{\beta \b t,\mu \b t}=Z^{-1} e^{\beta \hat{b}^{\dagger }\hat{b} \b \mu}~~,
\end{equation}
and consequently the system dynamics are described by a single non-linear differential equation
\begin{equation}
\dot{\beta}=k_{\downarrow}\b t\b{e^{\beta}-1}+k_{\uparrow}\b t\b{e^{-\beta}-1}~~,
\label{beta_dot}
\end{equation}
with initial conditions $\beta \b 0=-\f{\hbar \omega \b 0}{k_B T}$ and $\mu\b 0 =0$.
Equation \eqref{beta_dot} constitutes the basis for the suggested control scheme.

\paragraph*{Control}
The control target is to transform  a thermal state, defined by frequency $\omega_i$, to a thermal state of frequency $\omega_f$, while interacting with a bath at temperature $T$. The control utilizes the fact that at all times, the state is fully defined by $\mu\b t$ and $\beta \b t$. This property implies $\beta \b 0=-\f{\hbar \omega_i}{k_B T}$, $\beta \b{t_f}=-\f{\hbar \omega_f}{k_B T}$ and $\mu\b 0 =\mu\b{t_f}=0$.
The initial and final $\beta$ are connected through Eq. \eqref{beta_dot}, where the protocol defines the rates $k_\ua\b t$ and $k_\da\b t$. These rates are determined by the parameter $\alpha \b t$ in Eq. \eqref{eq:k down}, which in turn is completely defined by the control parameter $\omega \b t$ in Eq. \eqref{eq:alpha1}. Furthermore, $\mu \b{t}$ is determined by $\omega \b t$, and therefore $\omega \b t$ fully determines the state of the system at all times.

The strategy to solve the control equation is based on a reverse-engineering approach, and the protocol is denoted by {\em{Shortcut To Equilibration}} (STE).
The method proceeds as follows: we define a new variable $y=e^{\beta}$, and propose an ansatz for $y$ that satisfies the boundary conditions. Then we solve for  $\alpha\b t$, and from $\alpha \b t$ determine $\omega \b t$. 

The initial and final thermal states determine the boundary conditions of $\mu\b t$, which implies that the state is stationary at initial and final times. This leads to additional boundary conditions $\dot{\beta}\b 0=\dot{\beta}\b{t_f}=0$.

A third-degree polynomial is sufficient to obey all of the constraints. Introducing $s=t/t_f$, the solution reads
\begin{equation}
y\b{s}=y\b 0 +3 \Delta s^2 -2\Delta s^3~~,
\label{ap:y_s}
\end{equation}
where $\Delta=y\b{t_f}-y\b{0}$.
In principle, more complicated solutions for Eq. \eqref{beta_dot} exist; however, here we restrict the analysis to a polynomial solution \footnote{ Namely, the equation is of the Riccati form \cite{hazewinkel2013encyclopaedia,reid1972riccati}, and an analytical solution in an integral form  can be found, once the protocol is defined}. 
The implicit equation for $\alpha \b t$ becomes
\begin{multline}
    t_f\f{d}{ds}y\b s=k_{\downarrow}\b{\alpha\b s}y\b{s}^{2}-\\
    y\b{s}\b{k_{\downarrow}\b{\alpha\b s} +k_{\uparrow}\b{\alpha\b s}}+k_{\uparrow}\b{\alpha\b s}~~.
    \label{eq:y_dot}
\end{multline}
Solving the equation by numerical means generates  $\alpha\b s$. This solution is substituted into 
Eq. \eqref{eq:alpha1} and the control $\omega\b t$ is obtained by an iterative numerical procedure. The protocol satisfies the inertial condition on $\mu$, inferring that the derivation is self-consistent.

The solution of the STE incorporates the adiabatic result in the limit of slow driving. For large protocol time duration ($t_f \ra \infty$), the system's instantaneous state is a thermal state at temperature $T$ with frequency $\omega \b t$, see SM IV.   

\begin{figure}[htb!]
\centering
\includegraphics[width=0.49\textwidth]{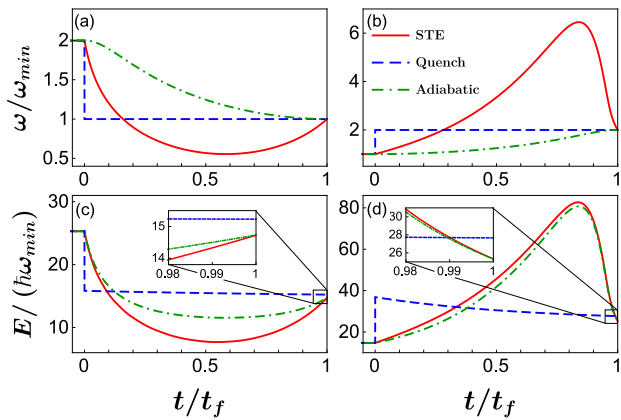}
\caption{Control protocols as a function of the scaled time $t/t_f$: (a,b) the oscillator frequency $\omega$ and (c,d) energy for the STE (red line), {\em{quench}} (dashed blue line) and adiabatic  (dot-dashed green line)
protocols. 
(a,c) Expansion, (b,d) compression  protocols.
The dynamics of the STE and quenched systems are shown for $t_f=8\, \text{a.u}$, and the adiabatic dynamics are obtained in the limit  $t_f\ra \infty$. (c,d) Inset: details of the final approach to the target state. Model parameters (atomic units): $\omega\b{0}/\omega_f=5/10$ for the compression, and reverse for the expansion and bath temperature $T=2$.} 
\label{fig:omega}
\end{figure}

We compare the STE protocol to a {\em{quench}} protocol involving a sudden change from $\omega \b 0=\omega_i$ to $\omega\b{t_f}=\omega_f$ \cite{de2010quench}. Two cases are studied, a compression of the potential, which corresponds to the transition $\omega\b 0 =5\ra \omega \b{t_f}=10$,  and a reversed expansion, associated with the transition $\omega\b 0 =10\ra \omega \b{t_f}=5$. Both protocols for each process are presented in Fig. \ref{fig:omega} panels (a) and (b). We add, as a reference, an adiabatic process obtained in the limit $t_f\ra \infty$.
  The initial stage of the {\it{quench}} protocol is effectively isolated, as the change in frequency is rapid relative to the relaxation rate toward equilibrium. As a result, the state stays constant while the Hamiltonian abruptly transforms to $\hat{H}_S \b{t_f}$. Coherence is generated with respect to $\hat{H}\b{t_f}$, because $\sb{\hat{H}\b{0},\hat{H}\b{t_f}}\neq 0$.
 After the initial stage energy is exchanged with the bath and the coherence dissipates.

% \begin{figure}[htb!]
%\centering
%\includegraphics[width=0.238\textwidth]{protocol_expansion2.eps}
%\includegraphics[width=0.238\textwidth]{protocol_compression2a.eps}
%\caption{Control protocol: The oscillator frequency $\omega$ as a function of time for a protocol time duration $t_f=8$, for expansion (left) and compression (right)  protocols. The parameters are in atomic units.} 
%\label{fig:omega}
%\end{figure}
 %The STE protocol modifies the trap potential, maintaining the inertial condition on $\mu$,.

\begin{figure}[htb!]
\centering
\includegraphics[width=0.49\textwidth]{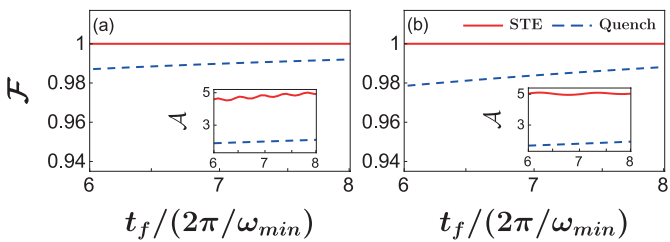}
\caption{The fidelity of the final state relative to the target thermal state for  the {\em{short-cut to equilibration}} (red) and  {\em{quench}} (blue) protocols. (a) Expansion protocol, (b) compression protocol. The inset shows the accuracy ${\cal{A}}=-\text{log}_{10}\b{1-{\cal{F}}}$, highlighting the 3-digit accuracy of the STE protocol. Model parameters are the same as in Fig. \ref{fig:omega}}\label{fig:fid} 
\end{figure}

In figure \ref{fig:fid}, we compare the fidelity with respect to the target thermal state of the expansion and compression protocols, for increasing stage times $t_f$ . The STE protocol transfers the system to the target thermal state with fidelities close to unity ${\cal{F}}\approx1$, while the {\em{quench}} target has lower fidelity due to the  slow relaxation. Therefore, the STE protocol equilibrates the system faster and with higher accuracy than the {\em{quench}} protocol. 
For a given fidelity, the STE achieves the target state up to five times faster than the {\em{quench}} protocol.

 %\begin{figure}[htb!]
%\centering
%\includegraphics[width=0.238\textwidth]{energy_expansion2b.eps}
%\includegraphics[width=0.238\textwidth]{energy_compression2.eps}
%\caption{Energy as a function of the scaled time $t/t_f$ for the STE (blue line), {\em{quench}} (dashed red line) and adiabatic (dot-dashed green line) protocols. The dynamics of the STE and quenched systems are shown for $t_f=8 \sb{\text{a.u}}$, and the adiabatic dynamics are obtained in the limit  $t_f\ra \infty$. Inset: details of the final approach to the target state.  Model parameters are the same as in Fig. \ref{fig:fid}.} 
%\label{fig:energy}
%\end{figure}
Figure \ref{fig:omega} panels (c) and (d) presents a comparison of the quantum state's energy for  the STE, {\em{quench}} and adiabatic protocols. %\tg{The STE and {\em{quench}} procedure have the same time duration, while the adiabatic process is given for $\tau_f\ra\infty$.}
During the {\em{quench}} protocol, there is a sudden change in the energy, which is followed by a slow exponential decay toward the thermal energy. The adiabatic and STE protocols are characterized by an overshoot  beyond the final thermal energy. In the final stage of the STE protocol, the energy rapidly converges to the desired thermal energy, whereas the quenched system remains far from equilibrium (see insets in Fig. \ref{fig:omega} panels (c) and (d)).

\paragraph*{Energy and entropy cost}
A control task can be evaluated by the work and entropy cost required to implement the control.  Restrictions on the cost can be connected to quantum friction \cite{feldmann2003quantum,plastina2014irreversible}, which implies that quicker transformations are accompanied by a higher energy cost \cite{chen2010transient,hoffmann2011time,salamon2009maximum,campbell2017trade,stefanatos2017minimum}.  Moreover, in any externally controlled process there is an additional cost in energy and entropy to generate faster driving \cite{torrontegui2017consumption,tobalina2018consumption}. The work cost for the STE protocol with a duration time $t$ is defined by the integral form
\begin{equation}
W \b t=\int_{0}^{t}{\text{tr}}\b{\hat{\rho}_S \b{t'}\pd{\hat{H}\b{t'}}{t'}}dt'~~.
\end{equation} 
For the {\em{quench}} protocol, the sudden transition occurs on a much faster timescale than the exchange rate of energy with the bath. This implies that the change in internal energy is equal to the work cost.
 \begin{figure}[htb!]
\centering
\includegraphics[width=0.3\textwidth]{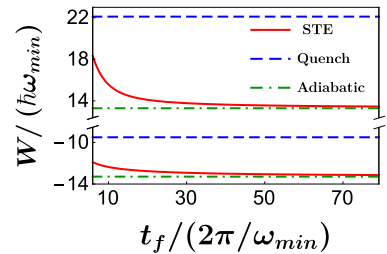}
\caption{Work required to perform the driving protocol as a function of the normalized time. Model parameters are identical to Fig. \ref{fig:fid}.
Upper part: compression, lower part: expansion.}
\label{fig:cost}
\end{figure}
For the expansion stroke (Fig. \ref{fig:cost}) the work generated during the STE protocol exceeds the quenched system result, yet remains below the adiabatic limit. When the system is compressed, the STE and {\em{quench}} protocols require additional work compared to the adiabatic process. We can define the efficiency of the process relative to the adiabatic work $W_{adi}$ ($\eta_{comp}=W_{adi}/W$ for compression and $\eta_{exp}=W/W_{adi}$ for the expansion). For the studied case, the efficiency of the STE protocol exceeds that of the {\em{quench}}, $\eta^{quench}_{comp}\approx0.6$, $\eta^{quench}_{exp}\approx0.7$, while $\eta^{STE}_{comp} >0.9$ and $\eta^{STE}_{exp} >0.75$, and improves for increasing protocol duration. This result is in accordance with thermodynamic principles, as any rapid driving will induce irreversible dynamics, which in turn leads to  sub-optimal performance.
For long times, the work of the STE procedure approaches the adiabatic result according to a $t^{-1}$ scaling law. At this limit, the global entropy production approaches zero. 
For shorter times, the system entropy change, for the STE procedure, is almost independent of protocol duration as a result of the accurate control. The price for shorter protocols is an increase in irreversibility, manifested by larger global entropy production (see
SM V).

  \paragraph*{Discussion}
  Quantum control is achieved by manipulating the system Hamiltonian via a change of an external control parameter. In turn, the change in the system Hamiltonian influences the system-bath interaction and the equation of motion. Hence, manipulating the Hamiltonian indirectly controls the dissipation rate.
  
  The control procedure employs a closed Lie algebra of system operators. The algebra is used to describe the Hamiltonian, system-bath interaction term and the state. The state is described by a generalized canonical form, Eq. \eqref{eq:Gibbs state}; this state is the maximum entropy state constrained by the expectation values of the operators in the algebra. For moderate acceleration of the driving, the {\em{inertial theorem}} can be employed to obtain the non-adiabatic master equation,  Eq. \eqref{eq:NAME} \cite{dann2018time,dann2018time}, for which the  generalized canonical form of the system state is preserved.

  Substituting the generalized canonical form in the equation of motion, Eq. \eqref{eq:NAME}, leads to a set of coupled non-linear differential equations of the state parameters, $\gamma$ and $\beta$, which define the generalized canonical state, Eq. \eqref{eq:Gibbs state}. These equations completely describe the system dynamics and implicitly depend on the control parameter. They are the basis for the control procedure. 

To solve the control problem, we insert a functional form for the state parameters which obeys the correct boundary conditions. Specifically, the parameters are associated with the initial and final thermal states, $\beta\b 0=-\hbar \omega_0/k_B T$ and $\beta\b{t_f}=-\hbar \omega_f/k_B T$, with vanishing derivatives at the boundaries.
 The considered functional form is a third-order polynomial, the coefficients of which are determined by the boundary conditions. This leads to an implicit equation in terms of the control parameter $\omega\b t$. 

At first glance, it would seem that the {\em{quench}} protocol is optimal, since the approach to equilibrium is exponentially fast. However, a superior solution is obtained by the STE protocol. The advantage of the latter is that it incorporates both the dissipative and unitary parts of the dynamics, changing the rates and engineering the state simultaneously.

 The related work cost, required for compression of the harmonic potential or obtained for expansion, is in accordance with thermodynamic principles. These infer that sudden or fast driving of the system increases the power output at the expense of wasted resources and entropy generation.

The STE protocol can be generalized beyond the isothermal example studied here, for three different kinds of scenarios: (i) the temperature of the initial state differs from the  bath temperature; (ii) the case of varying bath temperature (with the help of Eq. 
\eqref{eq:k down}); (iii) squeezed initial and final states. These general control tasks should be approached by reverse-engineering of both $\beta$ and $\gamma$ in Eq. \eqref{eq:beta_gamma}. Furthermore, once a non-adiabatic master equation is obtained \cite{dann2018time,dann2018inertial}, the method can be generalized to systems characterized by a closed Lie algebra.

 To conclude, the STE result demonstrates the feasibility of controlling the entropy of an open quantum system. Such control can be combined with fast unitary transformations to obtain a broad class of states within the system algebra. This will pave the way to faster high-precision quantum control, altering the state's entropy.  
 
 \paragraph*{Acknowledgement}
 We thank KITP for their hospitality, this research was supported in part by the Israel Science Foundation, grant number 2244/14, the National Science Foundation under Grant No. NSF PHY-1748958, the Basque Government, Grant No. IT986- 16 and  MINECO/FEDER,UE, Grant No. FIS2015-67161-P. We thank Marcel Fabian and J. Gonzalo Muga for fruitful discussions.

\appendix

%Substituting Eq. \eqref{ap:y_s} into Eq. \eqref{ap:y_dot} and solving numerically for  numerical solution 

%\bibliography{Equilibration2.bib}

\break
\onecolumngrid

\vspace*{0.4cm}
\begin{center}
	{\large \bf Shortcut to Equilibration of an Open Quantum System: supplementary material}
	\vskip 0.3 cm
	%{\large \bf} \\
	\vspace{0.2cm}
	Roie Dann,$^{1,2}$ Ander Tobalina,$^{3,2}$ Ronnie Kosloff,$^{1,2}$ \\
    \vspace{0.1cm}
	$^1${\small \it The Institute of Chemistry, The Hebrew University of Jerusalem, Jerusalem 9190401, Israel} \\
	$^2${\small \it Kavli Institute for Theoretical Physics, University of California, Santa Barbara, CA 93106, USA} \\
	$^3${\small \it Department of Physical Chemistry, University of the Basque Country UPV/EHU, Apdo 644, Bilbao, Spain}\\
	\vspace{0.1cm}
	%	$^{\color{blue}*}${\small hck@mail.ustc.edu.cn} \quad$^{\color{blue}\dagger}${\small ac\_santos@id.uff.br} \quad $^{\color{blue}\ddagger}${\small  jmcui@ustc.edu.cn} \quad $^{\color{blue}\S}${\small hyf@ustc.edu.cn} \quad  $^{\color{blue}\P}${\small dosp@ifsc.usp.br} \quad $^{\color{blue}**}${\small msarandy@id.uff.br} \quad $^{\color{blue}{\dagger\dagger}}${\small cfli@ustc.edu.cn} \quad $^{\color{blue}{\ddagger\ddagger}}${\small gcguo@ustc.edu.cn}
	
\end{center}
\vspace{0.6cm}

\twocolumngrid

\section{Canonical invarience and representation of the system in terms of the generalized Gibbs state coefficients.}
\label{ap:Gibbs}
We demonstrate how Lie algebra properties and canonical invariance can be utilized to obtain an alternative representation of the system dynamics. For a closed Lie algebra, the system state can be represented in a product form \cite{alhassid1978connection, Jaynes1957}. For the closed set ${\cal{B}}=\{\hat{b}^\dagger\hat{b},\hat{b}^2,\hat{b}^{\dagger2}\}$ the state is given by a generalized Gibbs state, presented in equation (5) of the Main Part (MP). Next, we calculate ${d\tilde{\rho}_S\b{t}}/dt\b{\tilde{\rho}_S\b{t}}^{-1}$ explicitly, using the generalized Gibbs state 
\begin{multline}
  \pd{\tilde{\rho}_{S}\b t}t\b{\tilde{\rho}_{S}\b t}^{-1}\\=  \dot{\gamma}\hat b^{2}+\dot{\hat \beta}e^{\gamma \hat b^{2}}\hat b^{\dagger}\hat be^{-\gamma \hat b^{2}}+\dot{\gamma}^{*}e^{\gamma \hat b^{2}}e^{\beta \hat b^{\dagger}\hat b}\hat b^{\dagger2}e^{-\beta \hat b^{\dagger}\hat b}e^{-\gamma \hat b^{2}}~~.
\end{multline}
Introducing the Baker–Campbell–Hausdorff relation, leads to
\begin{multline}
    \pd{\tilde{\rho}_{S}\b t}t\b{\tilde{\rho}_{S}\b t}^{-1}\\= \hat b^{2}\b{2\dot{\beta}\gamma+\dot{\gamma}+4\gamma^{2}\dot{\gamma}^{*}e^{2\beta}}+\hat b^{\dagger}\hat b\b{\dot{\beta}+4\gamma\dot{\gamma}^{*}e^{2\beta}}\\+\hat b^{\dagger2}\dot{\gamma}^{*}e^{2\beta}+\hat{I}2\gamma\dot{\gamma}^{*}e^{2\beta}~~.
    \label{eq:13}
\end{multline}
 By substituting Eq. (5) MP into the master equation, Eq. (2) MP, we obtain an expansion of  ${d\tilde{\rho}_S\b{t}}/dt\b{\tilde{\rho}_S\b{t}}^{-1}$ in terms of the operators of ${\cal{B}}$. Such a property is termed canonical invariance, it implies that an
initial state that belongs to the class of canonical states, will remain in this class throughout
the evolution \cite{rezek2006irreversible}. 
The expression reads
\begin{equation}
    \pd{\tilde{\rho}_{S}\b t}t\b{\tilde{\rho}_{S}\b t}^{-1}=\hat b^{2}A+\hat b^{\dagger}\hat bB+\hat b^{\dagger2}C+\hat I D~~,
    \label{eq:14}
\end{equation}
where
\begin{gather}
A={k}_{\downarrow}\b{a_{1}-\f 12a_{2}}-\f 12{k}_{\uparrow}a_{2}\\
  B={k}_{\downarrow}\b{b_{1}-\f 12b_{2}-\f 12}+{k}_{\uparrow}\b{b_{3}-\f 12b_{2}-\f 12}\\
  \nonumber
C=-\f 12\b{{k}_{\downarrow}+{k}_{\uparrow}}c_{1}+{k}_{\uparrow}c_{2}\\
\nonumber
D={k}_{\downarrow}\b{d_{1}-\f 12d_{2}}+{k}_{\uparrow}\b{-\f 12\b{d_{3}+1}}~~.\\
\nonumber
\end{gather}
The values of of the coefficients are summarized in Table \ref{tab:coeff}.
\begin{table}
\caption{\label{tab:coeff}}
\begin{ruledtabular}
\begin{tabular}{cccccccc}
 coefficient& value& coefficient & value &
\\
\hline
$a_1$& $2\gamma e^{\beta}$ & $c_1$ & $-2\gamma^{*}e^{2\beta}$ 
 \\ 
$a_2$ & $2\gamma\b{1-4|\gamma|^{2}e^{2\beta}}$  & $c_2$ & $-2\gamma^{*}e^{\beta}$ \\
$b_1$ & $e^{\beta}$ & $d_1$ & $e^{\beta}$   & \\
 $b_2$ & $1-8|\gamma|^{2}e^{2\beta}$ & $d_2$ & $-4|\gamma|^{2}e^{2\beta}$\\
 $b_3$ & $e^{-\beta}-4|\gamma|^{2}e^{\beta}$ & $d_3$ & $-2|\gamma|^{2}e^{\beta}+1$\\
\end{tabular}
\end{ruledtabular}
\end{table}
 To satisfy both Eq. \eqref{eq:13} and \eqref{eq:14} the coefficients multiplying each operator must be equal. Comparing terms, leads to four coupled differential equations
 \begin{multline}
     2\dot{\beta}\gamma+\dot{\gamma}+4\gamma^{2}\dot{\gamma}^{*}e^{2\beta}\\=\gamma\left(\left(2e^{\beta}-1\right)k_{\downarrow}+4e^{2\beta}\b{k_{\downarrow}+k_{\uparrow}}|\gamma|^{2}-k_{\uparrow}\right)
     \label{eq:set1}
 \end{multline}
 \begin{multline}
       \dot{\beta}+4\gamma\dot{\gamma}^{*}e^{2\beta}=e^{-\beta}\left(\left(e^{\beta}-1\right)\left(e^{\beta}k_{\downarrow}-k_{\uparrow}\right)\right.\\\left.+4e^{2\beta}|\gamma|^{2}\left(e^{\beta}\b{k_{\downarrow}+k_{\uparrow}}-k_{\uparrow}\right)\right)
       \label{eq:set2}
 \end{multline}
 \begin{equation}
       \dot{\gamma}^{*}e^{2\beta}=k_{\downarrow}e^{2\beta}\gamma^{*}+k_{\uparrow}\gamma^{*}\b{e^{2\beta}-2e^{\beta}}
        \label{eq:set3}
 \end{equation}
 \begin{multline}
   2\gamma\dot{\gamma}^{*}e^{2\beta}=k_{\downarrow}\b{e^{\beta}\left(2e^{\beta}|\gamma|^{2}+1\right)}\\+k_{\uparrow}\b{2e^{2\beta}|\gamma|^{2}-1} 
    \label{eq:set4}
 \end{multline}
After some algebraic manipulations we obtain the simplified form
\begin{gather}
\dot{\beta}=k_{\da}\b{e^\beta -1}+k_{\ua}\b{e^{-\beta}-1+4 e^{\beta}|\gamma|^2}\\
\nonumber \dot{\gamma}= \b{k_\da+k_\ua}\gamma-2 k_{\da} \gamma e^{-\beta}~~,
\end{gather}
These coupled differential equations completely determine the system's dynamics.

\section{Lower bound of the protocol duration}
\label{sec:lower bound}
The validity of the {\em{inertial theorem}} is quantified by the inertial parameter $\Upsilon$ \cite{dann2018inertial}. When $\Upsilon \ll1$ the inertial solution is a good approximation of the true dynamics. For the harmonic oscillator, $\Upsilon$ takes the form $\Upsilon \sim \f{\mu'\b{\theta}}{\b{2\kappa}^{2}}$, ($\theta=\theta\b t =\int_0^t \omega\b{t'}dt'$ and $\kappa = \sqrt{4-\mu^2}$), which explicitly becomes
 \begin{equation}
    \Upsilon= \f{1}{\b{2\kappa}^2}\b{ \f{\ddot{\omega}}{\omega^{3}}-2\mu^{2}}~~.
    \label{eq:Upsilon_HO}
 \end{equation}
 Transforming variables to the dimensionless parameter $s=t/t_f$ and constraining the inertial parameter by $\Upsilon<f\ll1$, introduces a lower bound for the protocol duration:
 \begin{equation}
t_{f}>f\cdot\text{max}_{s}\b{\f{1}{\omega}\sqrt{\f{\omega''\b s}{2\omega\b s}\f 1{8-\b{\mu\b s}^{2}}}}~~.     
 \end{equation}

\section{Quench protocol}
\label{Ap:quench}
When the parametric quantum harmonic oscillator is in a Gaussian state \cite{Jaynes1957}, it is convenient to analyze the system in terms of three time-dependent operators, the Hamiltonian Eq. (1) MP, Lagrangian $\hat{L}\b t=\f{\hat{P}^2}{2m}-\f{1}{2}m\omega^2 \b t \hat{Q}^2$, and the position-momentum correlation operator $\hat{C}\b t=\f{\omega\b t}{2}\b{\hat{Q}\hat{P}+\hat{P}\hat{Q}}$.  
The {\em{quench}} protocol includes an initial abrupt shift in frequency from $\omega \b 0=\omega_i$ to $\omega \b{t_f}=\omega_f$. The sudden transformation is approximately  isolated, as the bath's influence on the system occurs on a much longer timescale.
Moreover, for a sudden quench the system state remains unchanged. Hence, time-independent operators, such as $\hat{Q}\b 0$ and $\hat{P}\b 0$ do not vary. This property allows expressing the operators after the sudden quench $\hat{H}'$, $\hat{L}'$ and $\hat{C}'$ in terms of the operators at initial time (for the sudden quench) 
\begin{eqnarray}
\begin{array}{cc}
    \hat{H}'=\f{1}{2}\sb{\hat{H}\b 0\b{1+\f{\omega^2 \b t}{\omega^2 \b 0}}+\hat{L}\b 0\b{1-\f{\omega^2 \b t}{\omega^2 \b 0}}}\\
        \hat{L}'=\f{1}{2}\sb{\hat{H}\b 0\b{1-\f{\omega^2 \b t}{\omega^2 \b 0}}+\hat{L}\b 0\b{1+\f{\omega^2 \b t}{\omega^2 \b 0}}}\\
        \hat{C}'=\f{\omega \b t}{\omega \b 0}\hat{C}\b 0~~.
\end{array}
  \label{ap:sudden dyn}
\end{eqnarray}
The sudden change in frequency generates coherence, which is manifested by  non-vanishing values of $\mean{\hat{L}\b t}$ and $\mean{\hat{C}\b t}$ \cite{kosloff2017quantum}. 

Once the system is quenched, the frequency remains constant and the system relaxes towards equilibrium. Such dynamics were derived in Ref. \cite{kosloff2017quantum}, where the state's evolution is expressed as a matrix vector multiplication $\v v \b t ={\cal{U}}_S\b{t,0}\v v'$, with  $\v v \b t=\{\hat{H}\b t,\hat{L}\b t,\hat{C}\b t,\hat{I} \}$, $\v v'= \{ \hat{H}',\hat{L}',\hat{C}',\hat{I} \}$, $\hat{I}$ is the identity operator and 
\begin{equation}
    {\cal U}_{S}\b{t,0}=\sb{\begin{array}{cccc}
R & 0 & 0 & \mean{\hat H}_{eq}\b{1-R}\\
0 & Rc & -Rs & 0\\
0 & Rs & Rc & 0\\
0 & 0 & 0 & 1
\end{array}}~~.
\label{ap:propagator}
\end{equation}
Here, $R=e^{-\Gamma t}$ with $\Gamma = k_\da -k_\ua$,  $c=\cos\b{\omega_f t}$,  $s=\sin\b{\omega_f t}$ and $\mean{\hat H}_{eq}=\f{\hbar \omega_f}{2} coth\b{\f{\hbar \omega_f}{2 k_B T} }$.
Utilizing Eq. \eqref{ap:sudden dyn} and \eqref{ap:propagator}, the evolution of the quenched system is completely defined.

\subsection{Asymptotic behaviour of the fidelity of the quench procedure}
\label{sec:asymptotic fid}
The fidelity is a measure of the similarity between two quantum states. It was introduced by Uhlmann as the maximal
quantum-mechanical transition probability between the two states'
purifications in an enlarged Hilbert space \cite{uhlmann1976transition,jozsa1994r,marian2012uhlmann}. For two displaced squeezed thermal states the fidelity obtains the form \cite{scutaru1998fidelity,isar2009quantum}
\begin{equation}
    {\cal{F}}\b{\hat{\rho}_1,\hat{\rho}_2}=\f 2{\sqrt{\Delta+\delta}-\sqrt{\delta}}\exp\sb{-\v u^{T}\b{A_{1}+A_{2}}^{-1}\v u}~~,
\end{equation}
where $\Delta=\text{det}\b{A_{1}+A_{2}}$, $\delta=\b{\text{det}\b{A_{1}}-1}\b{\text{det}\b{A_{2}}-1}$, with 
\begin{equation}
    A_{i}=2\sb{\begin{array}{cc}
\sigma_{QQ}^{i} & \sigma_{PQ}^{i}/\hbar\\
\sigma_{PQ}^{i}/\hbar & \sigma_{PP}^{i}/\hbar^2
\end{array}}~~,
\end{equation}
where $\sigma_{QQ}$, $\sigma_{PP}$ and $\sigma_{PQ}$ are the variances and covariance of the position and momentum operators. The vector $\v{u}$ is given by, $\v u =\{\mean{\hat Q}_2 -\mean{\hat Q}_1,\mean{\hat P}_2 -\mean{\hat P}_1\}^T$. In the equilibration process the system's state remains centered at the origin (for all the considered protocols) and is compared to a thermal state. For such a case the calculation of the fidelity is greatly simplified, namely $\v u=0$ and the fidelity obtains the form,
\begin{equation}
    {\cal{F}}\b{\hat{\rho}_S\b{t_f},\hat{\rho}_{th}}=\f 2{\sqrt{\Delta+\delta}-\sqrt{\delta}}~~.
\end{equation}
For the {\em{quench}} procedure, once the sudden transition to the final frequency $\omega_f$ the system relaxes to a thermal state, following an exponential decay rate. For a time-independent Hamiltonian Eq. (2) MP, \cite{dann2018time}, reduces to the standard master equation for the harmonic oscillator  \cite{louisell1973quantum,lindblad1976brownian,breuer2002theory}:
\begin{multline}
\f d{dt}\hat{\rho}_{S} \b t= -i \omega\sb{\hat{a}^\dagger\hat{a},\hat{\rho}_S\b t}\\+ k_{\downarrow}\b{\hat{a}\rho_{S}\b t\hat{a}^{\dagger}-\f 12\{\hat{a}^{\dagger}\hat{a},\hat{\rho}_{S}\b t\}}\\
+k_{\uparrow} \b{\hat{a}^{\dagger}\hat{\rho}_{S}\b t\hat{a}-\f 12\{\hat{a}\hat{a}^{\dagger},\hat{\rho}_{S} \b t\}}~,
\label{eq:q optical master}
\end{multline}
where the annihilation operator is given by $\hat{a}=\b{\sqrt{\f{m\omega\b 0}{2\hbar}}\hat{Q}+\f{i}{\sqrt{2\hbar m\omega\b 0}}\hat{P}}$.
The solution of the master equation can be represented in the Heisenberg picture, obtaining the form \begin{gather}
    \hat{a}\b t=\hat{a}\b 0 e^{-i \omega -k/2 t}\\
    \hat{a}^{\dagger}\hat{a}\b t = \hat{a}^{\dagger} \hat{a}\b 0 e^{-k t}+ N\b{1-e^{-k t}}~~.
    \label{ap:static a dynamics}
\end{gather}
Next, we write the elements of $A_i$ in terms of the creation annihilation operators, and neglect terms that decay with a rate $2 k$. This leads to a simplified form for the fidelity
\begin{equation}
    {\cal{F}}\b{\hat{\rho}_S\b{t_f},\hat{\rho}_{th}}\approx\f 2{\sqrt{c_1+c_2 e^{-k t}}-\sqrt{c_3+c_4e^{-k t}}}~~,
    \label{eq:Fidelity 2}
\end{equation}
where $k\equiv k_{\da}-k_{\ua}$
\begin{gather}
    c_{1}=\b{4\sigma_{PP}^{th}\sigma_{QQ}^{th}/\hbar^{2}+1}{}^{2}\\
    \nonumber
    c_2 = 4\b{4\sigma_{PP}^{th}\sigma{}_{QQ}^{th}/\hbar^{2}+1}\b{c_{PP}\sigma_{QQ}^{th}+c_{QQ}\sigma_{PP}^{th}/\hbar^{2}}\\
    \nonumber
    c_3 = \b{4\sigma_{PP}^{th}\sigma{}_{QQ}^{th}/\hbar^{2}-1}^{2}\\
    \nonumber
    c_4 = 4\b{4\sigma_{PP}^{th}\sigma{}_{QQ}^{th}/\hbar^{2}-1}\b{\sigma_{PP}^{th}c_{QQ}/\hbar^{2}+\sigma{}_{QQ}^{th}c_{PP}}~~.
    \nonumber
\end{gather}
Here, $c_{PP}$ and $c_{QQ}$ are time-independent parameters, defined by the evolution of the system's variances
\begin{gather}
  \sigma_{QQ}=c_{QQ}e^{-k t}+\sigma_{QQ}^{th}~\\
  \nonumber
    \sigma_{PP}=c_{PP}e^{-k t}+\sigma_{PP}^{th}~.
\end{gather}
In the asymptotic limit, equation \eqref{eq:Fidelity 2} can be expanded in orders of $e^{-k t}$, and the fidelity's asymptotic behaviour reads: $ 1-{\cal{F}}\b t\propto e^{-k t}$.

\section{Adiabatic limit}
\label{subsec:adiabatic lim}
In the limit of infinite protocol time duration $t_f\ra \infty$ the STE result converges to the adiabatic solution. This can be seen by studying the change in $y\b t$. Differentiating  Eq. (9) MP leads to 
\begin{equation}
    \dot{y} \b t =6\Delta\f{t}{t_f^2}\b{1-\f{t}{t_f}}.
\end{equation}
Hence, in the limit $t_f\ra \infty$, $\dot{y}$ vanishes. Moreover, the effective frequency converges to $\alpha\b t\ra \omega \b t$ (Eq. (4) MP). 
Substituting this result into Eq. (10) MP gives the adiabatic solution
\begin{equation}
    y=\f{k_{\uparrow}^{adi}\b t}{k_{\downarrow}^{adi}\b t}~~,
    \label{ap:y adi}
\end{equation}
where the excitation and decay rate, in the adiabatic limit, are   $k_{\da}^{adi}\b t=k_{\ua}^{adi} \b t e^{\omega\b t/k_B T}=\f{\omega \b t|\v d|^{2}}{8\pi \varepsilon_{0} \hbar c}\b{1+N\b{\omega \b t}}$. Writing Eq. \eqref{ap:y adi} in terms of $\beta$, the system state (Eq. (5)) obtains the form
\begin{equation}
    \rho_S\b t=Z_{th}\b t e^{-\hbar\omega\b t \b{\hat{a}^{\dagger}\hat{a}+\f{1}{2}}},
\end{equation}
with $Z_{th}\b t=\text{tr}\b{ \exp\b{-\hbar\omega\b t\b{\hat{a}^{\dagger}\hat{a}+\f{1}{2}}}}$. Thus, in the adiabatic limit the STE solution converges to the adiabatic state.

 \begin{figure}[htb!]
\centering
\includegraphics[width=0.49\textwidth]{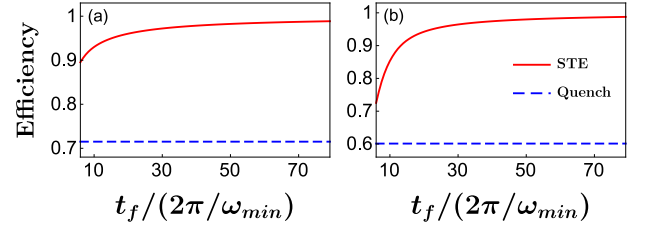}
\caption{The efficiency relative to the optimal adiabatic result for  the {\em{short-cut to equilibration}} (red) and  {\em{quench}} (blue) protocols, for (a) expansion protocol, (b) compression protocol.} 
\label{fig:Efficiency}
\end{figure}

\begin{figure}[htb!]
\centering
\includegraphics[width=0.49\textwidth]{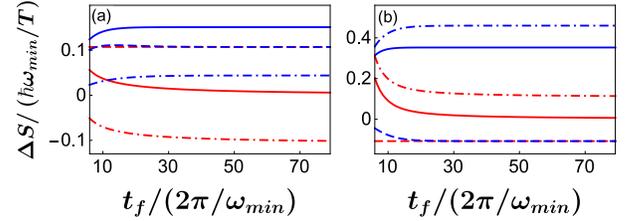}
\caption{The change in entropy during the (a) expansion and (b) compression processes for the STE (red) and {\em{quench}} (blue) protocols  as a function of protocol duration. The change in system entropy $\Delta S_{sys}$, bath entropy $\Delta S_B$ and global entropy $\Delta S_U$ are indicated by dashed, dashed-dot, and continuous lines respectively. } 
\label{fig:entropy}
\end{figure}

\section{Entropy calculation}
\label{sec:entropy}
The {\em{shortcut to equilibration}} procedure induces a swift change in the system's entropy $\Delta S_{sys}$. An expansion protocol is accompanied by an increase in the system entropy, while a compression is followed by a decrease in entropy,  Figure \ref{fig:entropy}. The STE transforms the system to the target state with high precision, which is almost independent on the protocol duration. As a result, the change in system entropy $\Delta S_{sys}$ also remains constant.

On the contrary, the global entropy generation depends on the trajectory between initial and final states. For large protocol times the STE approaches the adiabatic limit and the global entropy generation $\Delta S_U$ vanishes. The heat exchange with the bath increases for shorter protocol duration, which in turn, increases the change in the bath's entropy, $\Delta S_{B}=-Q_{sys}/T_B$. During the expansion (compression) process the system energy decreases (increases), and heat is transferred from (to) the bath, this is accompanied by and decrease (increase) in the bath entropy see Fig. \ref{fig:entropy}.

The change in entropy associated with the {\em{quench}} protocol is composed of a fast isoentropic process, followed by a natural decay towards equilibrium. During the relaxation stage the coherence decays, leading to a rise in the system entropy. For both expansion and compression quench protocols energy flows from the system to the bath, resulting in an increase in the bath entropy and a global entropy production. For both compression and expansion the global irreversible entropy generation of the {\em{quench}} exceeds the STE result.

 \section{Numerical details}
%$\omega\b 0/\omega\b{t_f}=5/10$ for the compression and reverse for the expansion, the bath temperature is $T=2$, coupling constant, $g=0.1$, $m=1$ $A=g^2\b{2/\hbar c}=g^2/2$
The value of $\omega\b t$ is assessed by a numerical solution of equations (10) MP and (4) MP, employing a built in Matlab solver. The solution is used to calculate the system's evolution according to the inertial solution \cite{dann2018inertial}. The validity of the inertial approximation has been verified, with the inertial parameter obtaining maximum values of $\Upsilon\approx0.1\,(0.04)$ for the compression (expansion) protocols.  \\
 The model parameters are summarized in Table \ref{tab:parameters}.
\begin{table}
\caption{\label{tab:parameters}}
\begin{ruledtabular}
\begin{tabular}{cccccccc}
 Coefficient& Value $\sb{\text{atomic units}}$ 
\\
\hline
Oscillator mass & $1$  \\
Compression $\omega\b{0}$ & 5 \\ Compression $\omega\b{t_f}$ & 10\\ 
Expansion $\omega\b{0}$ & 10 \\ Expansion $\omega\b{t_f}$ & 5 \\
Bath temperature & $2$ \\ 
Coupling prefactor &  $\f{|\v d|^{2}}{8\pi \varepsilon_{0} \hbar c}=0.02$    \\
\end{tabular}
\end{ruledtabular}
\end{table}

%\bibliography{main.bib}

\end{document}